\documentstyle[12pt,psfig]{article}
\def\reference{\parskip 0pt\par\noindent\hangindent 0.5 truecm}

\begin{document}
%
%
\title{Millimetric Astronomy from the High Antarctic Plateau: site
  testing at Dome C}
%


\author{Valenziano L. $^{1}$ \and
Dall'Oglio G. $^{2}$
} 

\date{}
\maketitle

{\center
$^1$ CNR-Te.S.R.E., via P. Gobetti 101, Bologna, Italy, I-40129\\valenziano@tesre.bo.cnr.it\\[3mm]
$^2$ Dipartimento di Fisica, Universit\`a di Roma TRE, via della Vasca
Navale 84, Roma, I-00154\\dalloglio@amaldi.fis.uniroma3.it\\[3mm]
}

%
\begin{abstract} Preliminary site testing at Dome C (Antarctica) is
presented, using both Automatic Weather Station (AWS) meteorological data
(1986-1993) and Precipitable Water Vapor (PWV) measurements made by the
authors. A comparison with South Pole and other sites is made. The South Pole
is a well established astrophysical observing site, where extremely good
conditions are reported for a large fraction of time during the year. Dome
C, where Italy and France are building a new scientific station, is a
potential observing site in the millimetric and sub-millimetric range. AWS
are operating at both sites and they have been continuously monitoring
temperature, pressure, wind speed and direction for more than ten years.
Site testing instruments are already operating at the South Pole (AASTO,
Automated Astrophysical Site-Testing Observatory), while {\it light}
experiments have been running at Dome C (APACHE, Antarctic Plateau
Anisotropy CHasing Experiment) during summertime. A direct comparison
between the two sites is planned in the near future, using the AASTO. The
present analysis shows that the average wind speed is lower at Dome C
($\sim$1 m/s) than at the South Pole ($\sim$2 m/s), while temperature and
PWV are comparable.
\end{abstract}


\bigskip

%
%

\section{Introduction}

The high Antarctic Plateau is considered the best site on Earth for
astrophysical observations in the millimetric and sub-millimetric wavelength
ranges (Burton et al. 1994). In the last few years many astrophysical
experiments have been deployed on this continent (Ruhl et al. 1995; Viper
Home Page; Balm 1996; Storey, Ashley \& Burton 1996 1996; Valenziano et al. 1998) and large
telescope projects are presently being developed (Stark 1998). The main
astrophysical observing site, already operational, is the U.S.
Amundsen-Scott station at the South Pole (SP hereafter), where the CARA
(Center for Astrophysical Research in Antarctica) installed quite a large
laboratory. Data on the observing conditions at SP have already been 
reported in the
literature (Dragovan et al. 1990; Chamberlin \& Bally 1994; Chamberlin
1995; Chamberlin, Lane \& Stark 1997; Lane 1998), while sophisticated tests are
being performed by Australian researchers, using an automated instrument
(Storey, Ashley \& Burton 1996).

Italy has been deeply involved in astrophysical activities in Antarctica since
1987. The Italian Antarctic Program (PNRA - Programma Nazionale di Ricerche
in Antartide) established an astrophysical observatory (OASI, Osservatorio
Antartico Submillimetrico e Infrarosso, Dall'Oglio et al., 1992) at the
Italian station Baia Terra Nova
(74$^{\circ}$ 41' 36" S, 164$^{\circ}$ 05' 58" E). 
A 2.6 meter sub-millimetric telescope is operating there during the
southern summer season. Many facilities are available, including Nitrogen
and Helium liquefiers, mechanical workshops, electronic and cryogenic
laboratories. The site quality is comparable to mid-latitude mountain
observatories for millimetric (Dall'Oglio et al. 1988) and mid-infrared
observations (Valenziano 1996).

In 1994, Italy and France started a program for building a permanent
scientific station (Concordia) on the high antarctic plateau, at Dome C,
hereafter DC, (75$^{\circ}$ 06' 25" S, 123$^{\circ}$ 20' 44" E). Domes are
regions more elevated than the rest of the continent (DC is at 3280 m),
barring the Trans Antarctic Mountains. The highest is Dome A
(4100 m), potentially the best observing site on the planet, but it is very
difficult to reach.

In December 1995, a test experiment was run to directly compare the short
term atmospheric stability (the so-called {\it sky noise}) between DC and
the Italian station on the coast. Raw data show a reduction in rms noise of
factors of 3 and 10 at $\lambda$=2 mm and $\lambda$=1.2 mm respectively
(Dall'Oglio 1997). In December 1996 the APACHE96 (Antarctic Plateau
Anisotropy CHasing Experiment) (Valenziano et al., 1997,1998) was set up at
DC. Preliminary data analysis shows good atmospheric stability in terms of
sky-noise at mm wavelengths. Atmospheric PWV content was measured during
the latter mission.

The AASTO experiment, presently running at the SP, is planned to be moved
to DC in the next few years. It will allow a careful assessment of the
observational quality at DC over a wide wavelength range. However, some
interesting information can be obtained from meteorological data, taking
advantage of the good statistics based on data collected over eight years.

The AWS project, started in the early 80s, installs automatic units in
remote areas of the Antarctic continent. The main objective of this program
is to support meteorological research by unattended, low cost, data
collecting stations.  AWS units operate continuously throughout the year.
AWS data for all the stations are publicly available at the University of
Wisconsin - Madison. 
One AWS unit is located a few kilometers from the Concordia Station site.
Therefore, it shares the same atmospheric conditions as the planned
astrophysics observatory. Another unit is at the geographical SP (Clean
Air, elevation 2836 m). Therefore it is possible to compare the two sites
on the basis of homogeneous data.

The main goal of this paper is to present data on DC 
conditions and to compare them with those of well established observing
sites. They are useful for planning experiments and for stimulating 
a deeper exploration of this interesting and promising observing site. 

\section{Automatic Weather Stations}

AWS were developed by the Radio Science Laboratory at Stanford University.
The basic AWS units measure air temperature, wind speed and wind direction
at a nominal height of three meters above the surface, and air pressure at
the electronic enclosure at about 1.75  meters above the surface. The
height is only nominal, due to possible snow accumulation. Some AWS units
can also measure humidity and vertical air temperature difference, but
these sensors are not available at either DC or SP stations. Data
are transmitted to a NOAA (National Oceanic and Atmospheric Administration)
satellite and then stored at the University of Wisconsin (Keller et al.
1997). More details on the AWS program are available at
http://uwamrc.ssec.wisc.edu/aws/.

\section{AWS data analysis}

 \begin{figure}[hbt]
 \psfig{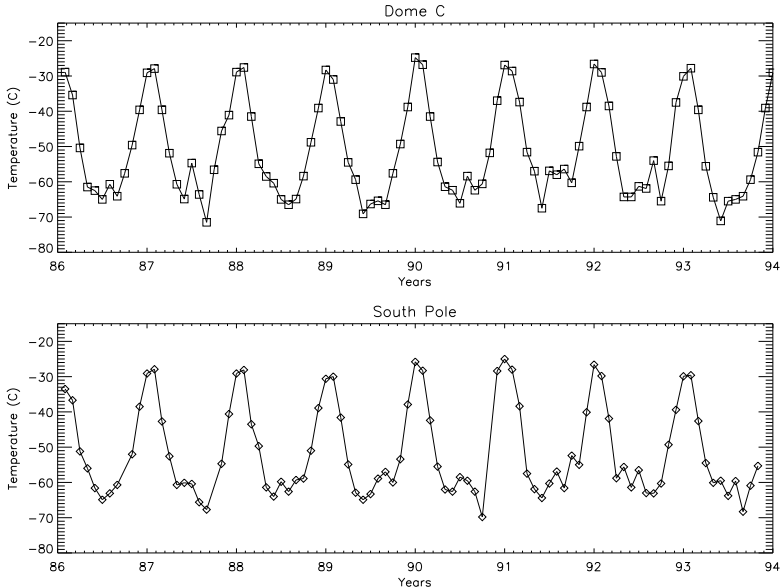}
 \caption{Monthly averaged temperature at DC (upper panel) and SP (lower
 panel). The correlation between the two data sets is 96\protect\%.}
 \label{month_t}            
\end{figure}

AWS data used in this work are already binned in  3 hour intervals. This
is useful in order to evaluate the stability of the observing conditions over
a reasonably short interval. Some data can be missing, due to instrumental
or transmission failures. Data were further averaged in one month
intervals. The typical uncertainty in the monthly averages is
15\% (standard deviation) for temperature, 1-6\% for pressure and
50-75\% for wind speed. Plots of these data are shown in Figures
\ref{month_t},\ref{month_p} and \ref{month_ws}. Statistical distributions of
the whole data set have also been calculated. Data have been binned in
1$^{\circ}$ intervals for temperature, 1 hPa for pressure, 1 m/s for wind
speed and 10$^{\circ}$ for wind azimuth. Histograms for temperature,
pressure, wind azimuth and wind speed are presented in Figures
\ref{month_t_dis},\ref{month_p_dis},\ref{wdir_dis} and \ref{cumul}. Monthly
plots of the whole data set and data distributions, along with a table with
median values and mean absolute deviation, are reported elsewhere
(Valenziano 1997).

\begin{figure}[hbtp]
 \psfig{file=t_cumul.plt,angle=90,height=8.5cm}
 \caption{Temperature distributions for DC (continuous line) and SP
   (dotted line). The two
   distributions are consistent within errors. Median temperature is
   -53$^{\circ}$ C. Data considered, over eight years, are 22596 for
   DC and 20814 for SP.}
 \label{month_t_dis}
   \psfig{file=monthly_p.plt,angle=90,height=8.5cm}
\caption{Average pressure for DC (upper panel) and SP (lower 
panel). The correlation between the two data sets is 92\%.}
 \label{month_p}            
\end{figure}
 
\begin{figure}[p]
 \psfig{file=p_cumul.plt,angle=90,height=8.5cm}
 \caption{Pressure distributions for DC (continuous line) and SP
   (dotted line). Median values are 644
   hPa (22600 data) and 682 hPa (20808 data), respectively. }
 \label{month_p_dis}
 \psfig{file=monthly_ws.plt,angle=90,height=8.5cm}
\caption{Average wind speed for DC (upper panel) and SP (lower 
panel). Correlation between data sets is less than 30\%.}
 \label{month_ws}            
\end{figure}
\begin{figure}
 \psfig{file=wd_cumul.plt,angle=90,height=8.5cm}
 \caption{Wind azimuth distributions for DC (continuous line) and SP
   (dotted line). The distribution
   for DC (23034 points) is peaked around azimuth 180, while SP 
   data (20784) are more uniformly distributed between azimuth 0
   and azimuth 100.}
 \label{wdir_dis}

 \psfig{file=ws_cumul.plt,angle=90,height=8.5cm}
 \caption{Histogram of data, with cumulative distribution of wind speed
   over-plotted, for DC (continuous line) and SP (dashed line). Number 
of data considered is 23038 and 20782, respectively. 
Distributions are not symmetric. The 50th
   percentile (0.5 cumulative probability of measuring lower values)
    is 1 m/s for DC and 2 m/s  for SP.}
\label{cumul}
 \end{figure}

\section{Precipitable Water Vapor measurements}

\begin{figure}[htb]
 \psfig{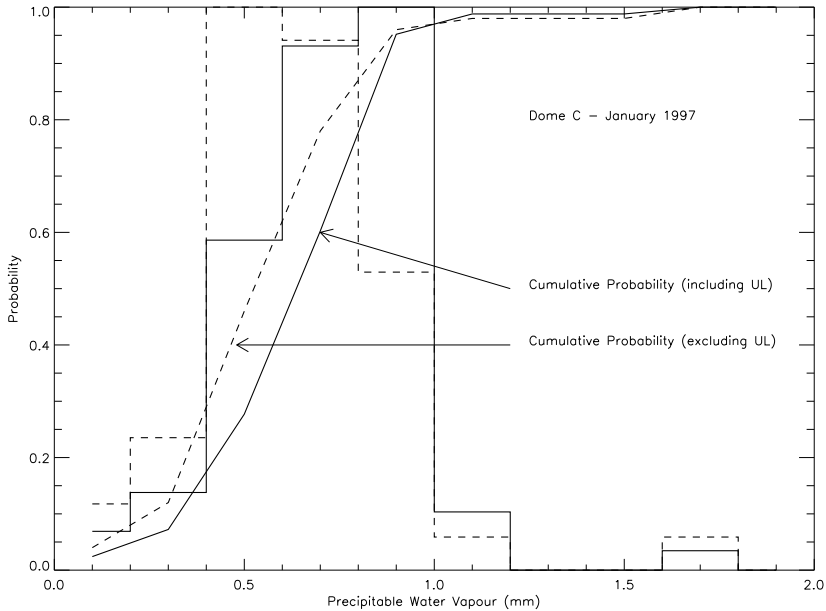}
 \caption{Precipitable Water Vapor data measured at DC during
 January 1997. The continuous line refers to the whole data set (83 measurements),
 including upper limits. Dashed line histogram is calculated 
 excluding upper limits (50 measurements). Cumulative probabilities are
 also plotted, with the same line styles.}
 \label{PWV}            
 \end{figure}

To the best of our knowledge, the first published measurement of the DC PWV
content of the atmosphere was performed by the authors in January 1997
(Valenziano et al. 1998). The instrument used was a portable photometer (Volz
1974), with an accuracy of 20 \%. The limited sensitivity of the instrument
allowed only upper limits to be set in some cases. Data are presented in
Figure \ref{PWV}. The average PWV at DC is around 0.6 mm.

A comparison between DC, SP, Atacama and Mauna Kea sites is reported in
Table \ref{tabpwv}. Data for DC were measured by the authors (in one
month), while quartile data for the other site are from Lane (1998).
Results for the Vostok station (elevation 3488 m) (Townes \& Melnick 1990) in
summer do not differ significantly from those reported here, but values of
less than 0.1 mm were measured during winter. We have also calculated the
225 GHz and 492 GHz opacities from the 50th percentile PWV data, using a
model evaluated from the SP data (Lane 1998). We used the following relations:
\begin{eqnarray}
\tau_0 (225 GHz) = 0.030 + 0.069 PWV (mm) \\
\tau_0 (492 GHz) = 0.33 + 1.49 PWV (mm)
\end{eqnarray}
\noindent and we evaluated the corresponding transmissions at the zenith 
as $T = \exp^{- \tau_0}$. These values are reported in Table
\ref{tabpwv}, along with available data for other sites (from Lane,
1998). 
\begin{table}[htbp]
\caption{Quartile of PWV, 225 GHz and 492 GHz opacity and
  transmissions at DC, SP, Mauna Kea and   Atacama. }
\bigskip

\begin{tabular*}{\textwidth}{lccccccc}
\hline
Site&\multicolumn{3}{c}{PWV (mm)}& $ \tau_0 $ & Transm. & $\tau_0 $&Transm. \\
&  25 \% & 50 \% & 75 \% & (225GHz) & (225GHz) &  (492GHz) & (492GHz)\\
\hline
Dome C (excluding UL)  & 0.38 & 0.52 & 0.68 & 0.066$^{1}$ & 0.94 
                       & 1.10$^{1}$ & 0.33\\
Dome C (including. UL) & 0.47 & 0.64 & 0.78 & 0.074$^{1}$ & 0.93 
                       & 1.28$^{1}$ & 0.28\\
South Pole (winter)    & 0.19 & 0.25 & 0.32 & 0.046 & 0.95 & 0.70 & 0.50\\
South Pole (summer)    & 0.34 & 0.47 & 0.67 & 0.062 & 0.94 & 1.03 & 0.36\\
Mauna Kea (winter)     & 1.05 & 1.65 & 3.15 & 0.076 & 0.93 & 1.32 & 0.28\\
Mauna Kea (summer)     & 1.73 & 2.98 & 5.88 & 0.129 & 0.88 & 2.38 & 0.09\\
Atacama (winter)       & 0.68$^{2}$ & 1.00$^{2}$ & 1.60$^{2}$ & 0.044 
                       & 0.96 & &\\
Atacama (summer)       & 1.10$^{2}$ & 2.00$^{2}$ & 3.70$^{2}$ & 0.077 
                       & 0.93 & &\\
\hline
    \end{tabular*}
    \label{tabpwv}
Data for sites other
than DC are from Lane, 1998. Values in   the first line for DC are
calculated for measurements only, while   those in the second line include
upper limits. Values for the zenith opacity,   referred to 50\% quartile,
for SP, Mauna Kea and Atacama are    measured with sky-dips. DC data are
calculated from 50\% quartile PWV.\\
$^{1}$From PWV values, using a model (see text).
$^{2}$PWV values are calculated from 225 GHz opacity (Lane 1998).
\end{table}

\section{Discussion}

The analysis of the AWS data set and the PWV measurements shows the 
following main results:
\begin{itemize}

\item DC and SP average temperatures are comparable, ranging 
between typical values of -65$^{\circ}$ C in winter and -26$^{\circ}$
C in summer. The median value for both sites is -53$^{\circ}$ C, while
the correlation of monthly average temperatures between DC and SP is 
  96\%.
\item The pressure is always lower at DC than at the SP. 
Median values are 644 hPa and 682 hPa respectively.
Monthly averaged pressure data show a correlation of 92 \% between the
  two sites.
\item  The wind speed is very low at both sites, with a maximum speed  of
  15.9 m/s at DC and 18.9 m/s at SP.
\item  A Kolmogorov-Smirnov test shows that wind speed distributions
  for the two sites are different. Correlation between monthly
  averaged wind speed data is less than 30 \%. 50th percentile 
  values for wind speed, evaluated for the whole data set, are 1 m/s 
  at DC and 2 m/s at SP.
\item Wind azimuth distributions are different: the prevalent direction is
  approximately azimuth 180 at DC and between azimuth 0 and 90 at SP.
\item PWV values measured at DC in January 1997 are comparable with SP
  values in the same season and lower than those measured at other sites.
\end{itemize}

In Table 1 and Table 2 our results for Antarctica (over eight years) are
compared with a well-established observing site, Mauna Kea (Hawaii Islands)
and a future important site in the Atacama desert (Northern Chile). Data for
these latter sites are reported from Holdaway (1996).

\begin{table}[htbp]
    \caption{Comparison between AWS data for Antarctic sites and data
      for Mauna Kea (Hawaii Islands) and Atacama desert (Chile) from 
Holdaway 1996}
    \begin{tabular*}{\textwidth}{l@{\extracolsep{\fill}}c@{\extracolsep{\tabcolsep}}ccc}
\hline
&Dome C&South Pole  &Mauna Kea  & Atacama  \\
\hline
Elevation (m) & 3280 & 2836 & 3750 & 5000 \\
Average pressure (hPa)  & 644 & 682 & 650 & 550\\
Median Wind speed (m/s)  & 2.1 & 3.6 & 4.5 & 6.0 \\
Maximum Wind Speed (m/s) & 15.9 & 18.5 & 28.8  & 33.0\\
\hline
    \end{tabular*}
    \label{tab_conf}
\end{table}

Some conclusions can be derived from these results:

\begin{itemize}
\item Lower wind regimes at the Antarctic sites result in lower
  turbulence implying a smaller contamination on observations. It is 
  worth considering that
  most of the {\it sky-noise} at Infra-Red and millimetric wavelengths
  is induced by convective motion and wind driven turbulence in the 
  lowest layers of the atmosphere, where the bulk of water vapor is 
  found (Smoot et al. 1987; Ade et al. 1984; Andreani et al 1990). 
  While the former needs
  further investigation (Argentini 1998, Burton 1995), the latter is
  minimal on the Antarctic Plateau.
\item Lower average wind speed reduces pointing errors for large
  antennas (see Holdaway, 1996).
\item In terms of wind speed, DC shows better conditions with respect
  to SP.
\item The high Antarctic Plateau is the driest observing site on
  Earth.
\item DC and SP pressure and temperature conditions are strongly
  correlated, indicating that they share similar meteorological
  conditions. It is possible to infer that PWV values at DC during
  wintertime are similar to SP ones in the same season.
\item 225 GHz and 492 GHz opacities for DC, calculated using a model
  valid for SP, show results similar to this site and better
  than Mauna Kea, when comparable water vapor amounts are considered. 
\end{itemize}

\section{Conclusions}

An analysis of available site quality data for DC has been accomplished,
including PWV measurements made by the authors. A comparison with SP and
other sites has been performed.

AWS data show that the quality of DC and SP is comparable, at least for
meteorological conditions. DC seems to be a more suitable site in terms of
wind speed (which is related to atmospheric turbulence).

The PWV content, measured at DC during the 1996-97 antarctic summer, is
comparable to that at SP during summertime. The high Antarctic Plateau is
shown to be the driest site compared to other observing sites. Further
measurements at DC, with improved sensitivity and automatic operation of
the instrument, are mandatory in order to explore conditions also
during wintertime.

In conclusion, the overall exceptional quality of the high Antarctic
Plateau at millimetric wavelengths, widely discussed in the literature
(Dragovan et al. 1990; Chamberlin \& Bally 1994; Burton et al. 1994; Chamberlin
1995; Chamberlin, Lane \& Stark 1997; Lane 1998, Stark 1998), is confirmed
by our work. DC conditions are at least comparable and probably better than
those at the SP. However, further and more sophisticated site testing
experiments are required to completely assess the DC observing quality.

\section*{Acknowledgments}

\noindent Data were provided by the Automatic Weather Station Project, run
by Dr. Charles R. Stearns at the University of Wisconsin - Madison, which
is funded by the National Science Foundation of the United States of
America. Data are available by anonymous ftp at ice.ssec.wisc.edu. This
work is partly supported by PNRA. We wish to thank G. Pizzichini, E. Pian
and J. Stephen for the careful revision of the text.

\section*{References}





\reference Ade, P.A.R. 1984, Infr. Phys., 24, 403
\reference Andreani P. et al., Infr. Phys., 30, 479
\reference Argentini S. 1998 in Atti della Conferenza Nazionale
sull'Antartide, (Roma: PNRA) 
\reference Balm, S.P. 1996 PASA, 13, 1, 14
\reference Burton, M. et al. 1994 PASA, 11, 127
\reference Burton, M.G. 1995, in ASP Conf. Ser. 73, Airborne
     Astronomy Symp. on the Galactic Ecosystem: From Gas to Stars to
     Dust, eds. M. R. Haas, J. A. Davidson \& E. F. Erickson, 
     (San Francisco: ASP), 559-562.
\reference Chamberlin, R.A.  1995 Int. J. Infr. Millim. 
Waves, 16, 907
\reference Chamberlin, R.A.\& Bally, J. 1994 Appl. Opt, 
33, 1095
\reference Chamberlin, R.A., Lane, A.P. \& Stark, A.A. 1997 Ap.J., 476, 428 
\reference Dall'Oglio, G. 1997 Communication at the Concordia Project 
Meeting, Siena (Italy), June 3-6
\reference Dall'Oglio, G., et al. 1988 Infr. Phys., 28, 155
\reference Dall'Oglio, G., et al. 1992 Exp. Astr., 2, 275
\reference Dragovan, M., Stark, A. A., Pernic, R. \& Pomerantz,
M. A. 1990, Appl. Opt., 29, 463 
\reference Gundersen, J.O., et al. 1995 Ap.J., 443, L57
\reference Holdaway, M.A., et al., 1996, MMA Memo 159
\reference Keller, L.M., et al. 1997 Antarctic AWS data for Calendar Year 
1995, Space Science and Engineering Center, Univ. of Wisconsin, Madison 
USA
\reference Lane, A.P. 1998, in ASP Conf. Ser. 141, Astrophysics from
Antarctica, ed. R. Landsberg \& G. Novak (San Francisco: ASP), 289
\reference Ruhl, J.R., et al. 1995 Ap.J., 451, L1
\reference Smoot, G.F. et al. 1987, Radio Sci., 22, 521-528
\reference Stark, A. 1998, in ASP Conf. Ser. 141, Astrophysics from
Antarctica, ed. R. Landsberg \& G. Novak (San Francisco: ASP), 349  
\reference Storey, J.W.V., Ashley, M.C.B., Burton, M.G. 1996 PASA, 13, 35
\reference Townes C.H. \& Melnick, G. 1990 PASP, 102, 357
\reference Valenziano, L. 1996 Ph.D. thesis, Universit\`a
di Perugia
\reference Valenziano, L. 1997 Te.S.R.E. Rep. 192/97, available at \\
http://tonno.tesre.bo.cnr.it/~valenzia/APACHE/apache.htm
\reference Valenziano, L., et al. 1997 Proc. PPEUC, available at\\
http://www.mrao.cam.ac.uk/ppeuc/astronomy/papers/valenziano/valenziano.html
\reference Valenziano, L., et al. 1998, in ASP Conf. Ser. 141, 
Astrophysics from Antarctica, ed. R. Landsberg \& G. Novak 
(San Francisco: ASP), 81
\reference Viper Home Page, http://cmbr.phys.cmu.edu/vip.html
\reference Volz, F. 1974 Appl. Opt., 13, 1732
\end{document}